\documentclass[twocolumn,showpacs,preprintnumbers,amsmath,amssymb]{revtex4}
\usepackage{amsfonts}
\usepackage{mathrsfs}
\usepackage{graphicx}
\usepackage{dcolumn}
\usepackage{bm}

\begin{document}

\title{The Dynamics of the Bounds of Squared Concurrence}

\author{Zhao Liu}%
 \email{liuzhaophys@aphy.iphy.ac.cn}
\author{Heng Fan}
 \email{hfan@aphy.iphy.ac.cn}
\affiliation{%
Institute of Physics, Chinese Academy of Sciences, Beijing 100190,
China
}%
\date{\today}

\begin{abstract}
The dynamics of the quantum entanglement is a fundamental
characteristic for various quantum systems. Since the computable
entanglement measure for higher dimensional quantum states itself is
absent, the dynamics of the entanglement expressed in an operational
method will be of interest. We study the dynamics of $\tau$, an
analytical lower bound of squared concurrence, of a bipartite
$\textit{d}\otimes\textit{d}$ quantum state when one party goes
through an arbitrary noisy channel. For a pure input state, the
range of $\tau$ is obtained explicitly. For a mixed input state, an
upper bound of $\tau$ is found. Interestingly, the tangle $\tau'$,
as an upper bound of squared concurrence, also has a similar
dynamical property. Our results are similar to that of Konrad
\emph{et al}. and can help the estimation of high-dimension
bipartite entanglement in experiments.
\end{abstract}

\maketitle


{\label{sec:level1}} \section{Introduction} Quantum entanglement,
which is considered to be the most non-classical phenomenon in the
quantum world, lies in the central position of quantum information
theory (QIT). It has been identified as a key resource in many
aspects of QIT, such as quantum teleportation, quantum key
distribution and quantum computation \footnote{M.A.Nielsen and
I.L.Chuang: \emph{Quantum Computation and Quantum information},
Cambridge University Press, Cambridge 2000.}. But while implementing
quantum information precessing in real physical systems, it's
inevitable that the entanglement decays due to the interactions of
our system with the environment, making it significantly important
to study the dynamical property of entanglement in realistic
situations.

The dynamical property, namely the time evolution of entanglement of
a state is usually deduced from the time evolution of the state
itself \footnote{D.Braun, Phys. Rev. Lett. \textbf{89}, 277901
(2002).}\footnote{J.P.Paz and A.J.Roncaglia, Phys. Rev. Lett.
\textbf{100}, 220401 (2008).}. However, recently, in
Ref.\footnote{T.Konrad, F.De Melo, M.Tiersch, C.Kasztelan, A.Aragao
and A. Buchleitner, Nature Physics \textbf{4}, 99 (2008).}, without
solving the master equation of a quantum state but by utilizing the
Jamiolkowski isomorphism, Konrad \emph{et al.} presented a
factorization law for a two qubit system, which describes the
evolution of entanglement in a simple and general way. Then, Li
\emph{et al.} generalized this result to that of a bipartite quantum
system of arbitrary dimension \footnote{Z.G.Li, S.M.Fei, Z.D.Wang
and W.M.Liu, arXiv: 0806.4228v3.}. In the study above, concurrence
which is a well accepted entanglement measure, was used to quantify
the entanglement. As is well known, for higher dimensional bipartite
quantum state, there is no analytic method in general to find
concurrence. Thus it will be very interesting if we can study the
dynamical of the entanglement which is quantified in an operational
way. Unfortunately, there is no such an operational measure of
entanglement for an arbitrary bipartite quantum state. However,
there exist a lower bound of squared concurrence which is analytic
\footnote{Y.C.Ou, H.Fan and S.M.Fei, Phys. Rev. A \textbf{78},
012311 (2008).} and we represent it as $\tau $. In this paper, we
will investigate the dynamical of the lower bound of the squared
concurrence $\tau$. As a special case for two-qubit state, our
result reduces to the result by Konrad \emph{et al}. Moreover, the
tangle $\tau'$ as defined in Ref.[7] is an upper bound of squared
concurrence. Interestingly, it has a similar dynamical property with
$\tau$. To clarify our results, we use the depolarizing and the
phase damping channels as the examples.


{\label{sec:level1}} \section{Concurrence and its upper and lower
bounds} As the beginning we recall the definition of concurrence,
$\tau$ and tangle $\tau'$. For a pure bipartite state
$\rho_{\emph{AB}}=\mid\psi\rangle\langle\psi|$ in a finite
$d_{1}\otimes d_{2}$ dimensional Hilbert space $\mathcal
{H}_A\otimes\mathcal {H}_B$, the concurrence is defined as $\mathcal
{C}(|\psi\rangle)=\sqrt{2(1-Tr\rho_A^2)}$, with
$\rho_A=Tr_B{\rho_{AB}}$ the reduced density matrix. For a mixed
bipartite state
$\rho=\sum_{i}p_{i}|\psi_{i}\rangle\langle\psi_{i}|,p_{i}\geq0,\sum_{i}p_{i}=1$,
the concurrence is defined as the convex roof of all possible
decompositions of $\rho$ into the pure states $|\psi_{i}\rangle$,
namely $\mathcal {C}(\rho)\equiv
\min_{\{p_{i},|\psi_{i}\rangle\}}\sum_{i}p_{i}\mathcal
{C}(|\psi_{i}\rangle)$.

Although the concurrence of a general bipartite mixed state defined
above is difficult to solve due to a high-dimensional optimization,
a computable lower bound of squared concurrence can be found in
Ref.[6]:
\begin{equation}
\mathcal
{C}^{2}(\rho)\geq\sum^{d_{1}-1}_{r>p=0}\sum^{d_{2}-1}_{r'>p'=0}\mathcal{C}^{2}_{pr,p'r'}(\rho)\equiv\tau(\rho),
\end{equation}
where $\tau$ is a lower bound of squared concurrence and
\begin{equation}
\mathcal {C}_{pr,p'r'}(\rho)=\max
\{0,\lambda^{1}_{pr,p'r'}-\lambda^{2}_{pr,p'r'}-\lambda^{3}_{pr,p'r'}-\lambda^{4}_{pr,p'r'}\},
\end{equation}
with $\lambda^{i}_{pr,p'r'}$ being the squared roots of the four
nonzero eigenvalues, in decreasing order, of
$\rho\tilde{\rho}_{pr,p'r'}$, where
$\tilde{\rho}_{pr,p'r'}=(L_{pr}\otimes
L_{p'r'})\rho^{*}(L_{pr}\otimes L_{p'r'})$ and
$L_{pr}=|p\rangle\langle r|-|r\rangle\langle
p|(p,r=0,1,...,d_{1}-1;p<r)$, $L_{p'r'}=|p'\rangle\langle
r'|-|r'\rangle\langle p'|(p',r'=0,1,...,d_{2}-1;p'<r')$ are the
generators of the group SO$(d_{1})$ and SO$(d_{2})$ respectively.
It's clear that $\tau$ can always be calculated analytically.

According to Ref.[6], every $\mathcal {C}_{pr,p'r'}(\rho)$ can be
seen as a two qubit concurrence of a $4\times4$ matrix $\tilde {\rho
}$, which is a submatrix of $\rho$,
\begin{equation}
\tilde {\rho }=\left(
\begin{array}{cccc}
\rho_{pp',pp'}&\rho_{pp',pr'}&\rho_{pp',rp'}&\rho_{pp',rr'}\\
\rho_{pr',pp'}&\rho_{pr',pr'}&\rho_{pr',rp'}&\rho_{pr',rr'}\\
\rho_{rp',pp'}&\rho_{rp',pr'}&\rho_{rp',rp'}&\rho_{rp',rr'}\\
\rho_{rr',pp'}&\rho_{rr',pr'}&\rho_{rr',rp'}&\rho_{rr',rr'}
\end{array}
\right)
\end{equation}
with subindices \emph{p} and \emph{r} associated with $\mathcal
{H}_{A}$ and $p'$ and $r'$ with $\mathcal {H}_{B}$. So $\tau(\rho)$
in fact is the sum of some two qubit entanglement in a high
dimensional state, according to which we can rewrite Eq.(1) in
another form:
\begin{equation}
\tau(\rho)=\sum_{i=1}^{\mathcal {D}}\mathcal {C}^{2}(\tilde {\rho
}_{i}),
\end{equation}
where $\mathcal {C}$ is just the two qubit concurrence and $\tilde
{\rho }_{i}$ is a submatrix of $\rho$ of the form (3), the number of
which is $\mathcal {D}=\frac{d_{1}d_{2}(d_{1}-1)(d_{2}-1)}{4}$.

One can prove that $\tau$ is a convex function of the density
operator. According to the definition,
$\tau\Big(\sum_{i}p_{i}\rho_{i}\Big)=\sum_{k=1}^{\mathcal
{D}}\mathcal {C}^{2}\Big(\sum_{i}p_{i}\tilde {\rho }_{i}^{k}\Big)$,
where $\tilde {\rho }_{i}^{k}$ is a submatrix of $\rho_{i}$. Using
the convexity of $\mathcal {C}$,
$\tau\Big(\sum_{i}p_{i}\rho_{i}\Big)\leq\sum_{k=1}^{\mathcal
{D}}\Big(\sum_{i}p_{i}\mathcal {C}(\tilde {\rho
}_{i}^{k})\Big)^{2}$. Recall that $f(x)=x^{2}$ is a convex function,
namely $(\sum_{i}p_{i}x_{i})^{2}\leq\sum_{i}p_{i}x^{2}_{i}$ for
$p_{i}\geq0,\sum_{i}p_{i}=1$. So
\begin{eqnarray}
\tau\Big(\sum_{i}p_{i}\rho_{i}\Big)\leq\sum_{k=1}^{\mathcal
{D}}\sum_{i}p_{i}\mathcal
{C}^{2}(\tilde {\rho }_{i}^{k})\nonumber\\
=\sum_{i}p_{i}\tau(\rho_{i}),
\end{eqnarray}
which is just what we want to prove.

The tangle for a general mixed state is defined as
\begin{equation}
\tau'(\rho)\equiv
\min_{\{p_{i},|\psi_{i}\rangle\}}\sum_{i}p_{i}\mathcal
{C}^{2}(|\psi_{i}\rangle),
\end{equation}
which is also a convex function of the density operator. One can
easily see that $\tau'(\rho)\geq\mathcal {C}^{2}(\rho)$ from the
convexity of concurrence so it's an upper bound of squared
concurrence.

{\label{sec:level1}} \section{The dynamics of concurrence} Here we
briefly review the dynamics of concurrence demonstrated in Ref.[4]
and Ref.[5]. First consider a $2\otimes2$ two qubit pure state,
after only one qubit goes through an arbitrary channel $\mathcal
{E}$, the concurrence between them decays just by a universal factor
only determined by $\mathcal{E}$'s action on the maximally entangled
state $|\phi^{+}\rangle=\frac{1}{\sqrt{2}}(|00\rangle+|11\rangle)$:
\begin{equation}
\mathcal {C}[({\bf{1}}\otimes\mathcal
{E})|\psi\rangle\langle\psi|]=\mathcal {C}[({\bf{1}}\otimes\mathcal
{E})|\phi^{+}\rangle\langle\phi^{+}|]\mathcal {C}(|\psi\rangle).
\end{equation}

For a general $d_{1}\otimes d_{2}$ pure state, a similar relation is
satisfied, with a sacrifice that the equality is replaced by an
inequality:
\begin{equation}
\mathcal {C}[({\bf{1}}\otimes\mathcal
{E})|\psi\rangle\langle\psi|]\leq\frac{d_{2}}{2}\mathcal
{C}[({\bf{1}}\otimes\mathcal
{E})|\phi^{+}\rangle\langle\phi^{+}|]\mathcal {C}(|\psi\rangle).
\end{equation}

Both results above can be generalized to the case where the input
state is a mixed state. For a $2\otimes2$ mixed state we have
\begin{equation}
\mathcal {C}[({\bf{1}}\otimes\mathcal {E})\rho_{0}]\leq\mathcal
{C}[({\bf{1}}\otimes\mathcal
{E})|\phi^{+}\rangle\langle\phi^{+}|]\mathcal {C}(\rho_{0})
\end{equation}
and for a $d_{1}\otimes d_{2}$ mixed state we have
\begin{equation}
\mathcal {C}[({\bf{1}}\otimes\mathcal
{E})\rho_{0}]\leq\frac{d_{2}}{2}\mathcal
{C}[({\bf{1}}\otimes\mathcal
{E})|\phi^{+}\rangle\langle\phi^{+}|]\mathcal {C}(\rho_{0}).
\end{equation}


{\label{sec:level1}} \section{The dynamics of $\tau$ and $\tau'$}
Generally speaking, to solve the concurrence of a high-dimensional
mixed state, just like $({\bf{1}}\otimes\mathcal
{E})|\phi^{+}\rangle\langle\phi^{+}|$, one must make an optimal
decomposition of the state, which is a notoriously difficult task,
making the right hand side of Eq.(8) and Eq.(10) nearly impossible
to be calculated analytically except in some special cases. This
motivates us to investigate the time evolution of $\tau$, which can
be calculated analytically.

Let us consider a $d\otimes d$ bipartite quantum system whose
Hilbert space is $\mathcal {H}$, then any pure state
$|\psi\rangle\in\mathcal {H}$ can be expressed by Schmidt
decomposition as follows:
\begin{equation}
|\psi\rangle=\sum^{d-1}_{i=0}\sqrt{\omega_{i}}|ii\rangle,\sum^{d-1}_{i=0}\omega_{i}=1
\end{equation}
and the maximally entangled state in $\mathcal {H}$ can be written
as $|\phi^{+}\rangle=\frac{1}{\sqrt{d}}\sum^{d-1}_{i=0}|ii\rangle$.

Because Jamiolkowski isomorphism can be extended to bipartite
systems of arbitrary finite dimension, the dual picture used in
Ref.[4] will be valid in higher dimensions. Consider a quantum
channel $\mathcal {E}$, according to Jamiolkowski isomorphism, when
only one qubit of the state (11) goes through $\mathcal {E}$, we
have $\rho'=\frac{({\bf{1}}\otimes\mathcal
{E})|\psi\rangle\langle\psi|}{p'}=\frac{(\mathcal
{E}_{\psi}\otimes{\bf{1}})\rho_{\mathcal {E}}}{p}$, where $
\rho_{\mathcal {E}}=\frac{({\bf{1}}\otimes\mathcal
{E})|\phi^{+}\rangle\langle\phi^{+}|}{p''}$ with $p$, $p'$ and $p''$
the normalization coefficients and one can verify that
$p'=d^{2}pp''$. The action of channel $\mathcal {E}_{\psi}$ can be
expressed in a simple form that $(\mathcal
{E}_{\psi}\otimes{\bf{1}})\rho_{\mathcal {E}}=(\mathcal
{M}\otimes{\bf{1}})\rho_{\mathcal {E}}(\mathcal
{M}^{\dagger}\otimes{\bf{1}})$, where $\mathcal
{M}=\frac{1}{\sqrt{d}}\sum^{d-1}_{i=0}\sqrt{\omega_{i}}|i\rangle\langle
i|$.

Because $\mathcal {M}^{\dagger}=\mathcal {M}$ and $\mathcal
{M}L_{pr}\mathcal {M}=\frac{\sqrt{\omega_{p}\omega_{r}}}{d}L_{pr}$,
$\det(\rho'\tilde{\rho'}_{pr,p'r'}-\lambda{\bf{1}})
=\det\Big(\frac{\omega_{p}\omega_{r}}{d^{2}p^{2}}\rho_{\mathcal
{E}}\tilde{\rho_{\mathcal {E}}}_{pr,p'r'}-\lambda{\bf{1}}\Big)$,
from which we can have that $\mathcal
{C}^{2}_{pr,p'r'}(({\bf{1}}\otimes\mathcal
{E})|\psi\rangle\langle\psi|)=d^{2}\omega_{p}\omega_{r}\mathcal
{C}^{2}_{pr,p'r'}(({\bf{1}}\otimes\mathcal
{E})|\phi^{+}\rangle\langle\phi^{+}|)$. Noting $\mathcal
{C}^{2}_{pr,p'r'}(|\psi\rangle)=4\omega_{p}\omega_{r}\delta_{pp'}\delta_{rr'}$,
an important relation is derived:
\begin{eqnarray}
\mathcal {C}^{2}_{pr,p'r'}(({\bf{1}}\otimes\mathcal
{E})|\psi\rangle\langle\psi|)=\frac{d^{2}}{4}\Big(\sum^{d-1}_{r''>p''=0}\mathcal{C}_{pr,p''r''}(|\psi\rangle)\nonumber\\
\times\mathcal {C}_{p''r'',p'r'}(({\bf{1}}\otimes\mathcal
{E})|\phi^{+}\rangle\langle\phi^{+}|)\Big)^{2}.
\end{eqnarray}

In \emph{Introduction} we have explained $\mathcal {C}_{pr,p'r'}$ as
a two qubit concurrence, so Eq.(12) means that the evolution of a
certain two qubit entanglement in a high dimensional state also
obeys a law which is similar to Eq.(7) but more complicated because
we must consider all related two qubit entanglement, as demonstrated
in the sum in the RHS of Eq.(12). It's easy to see that when $d=2$,
Eq.(12) is equivalent to Eq.(7).

In what follows we want to find the range of
$\tau(({\bf{1}}\otimes\mathcal {E})|\psi\rangle\langle\psi|)$.
According to the definition of $\tau$, we have
\begin{eqnarray}
\tau(({\bf{1}}\otimes\mathcal {E})|\psi\rangle\langle\psi|)=
\frac{d^{2}}{4}\mathcal {C}^{2}(|\psi\rangle)\sum^{d-1}_{r>p=0}
\sum^{d-1}_{r'>p'=0}\frac{\omega_{p}\omega_{r}}{\sum^{d-1}_{i<j=0}\omega_{i}\omega_{j}}
\nonumber \\
 \times\mathcal{C}^{2}_{pr,p'r'}(({\bf{1}}\otimes\mathcal
{E})|\phi^{+}\rangle\langle\phi^{+}|).\nonumber\\
\end{eqnarray}
Considering
$\omega_{p}\omega_{r}\leq\sum^{d-1}_{i<j=0}\omega_{i}\omega_{j}$, we
immediately get the upper bound of $\tau(({\bf{1}}\otimes\mathcal
{E})|\psi\rangle\langle\psi|)$:
\begin{equation}
\tau(({\bf{1}}\otimes\mathcal
{E})|\psi\rangle\langle\psi|)\leq\frac{d^{2}}{4}\tau(({\bf{1}}\otimes\mathcal
{E})|\phi^{+}\rangle\langle\phi^{+}|)\mathcal {C}^{2}(|\psi\rangle).
\end{equation}
On the other hand, one can show that
$\sum^{d-1}_{i<j=0}\omega_{i}\omega_{j}=\frac{1}{2}(1-\sum^{d-1}_{i=0}\omega_{i}^{2})\leq\frac{d-1}{2d}$.
Let $\eta=\min_{\{p,r\}}\omega_{p}\omega_{r}$ for any pair $p<r$
satisfying $\omega_{p}\omega_{r}\neq0$, we find a lower bound of
$\tau(({\bf{1}}\otimes\mathcal {E})|\psi\rangle\langle\psi|)$:
\begin{equation}
\tau(({\bf{1}}\otimes\mathcal
{E})|\psi\rangle\langle\psi|)\geq\frac{2d\eta}{d-1}\frac{d^{2}}{4}\tau(({\bf{1}}\otimes\mathcal
{E})|\phi^{+}\rangle\langle\phi^{+}|)\mathcal {C}^{2}(|\psi\rangle).
\end{equation}

Eq.(14) and Eq.(15) are our central results. Both of them have the
form of a factorization law similar to Eq.(7) and Eq.(8). In
Eq.(14), the factor is universal determined only by the channel's
action on the maximally entangled state. But in Eq.(15), the factor
includes $\eta$ relevant to the input state itself, which, however,
is easy to compute by contrast to the evolution of the input state.
So in order to know the dynamics of $\tau$ of some pure input
states, we only need to study the dynamics of $\tau$ of the
maximally entangled state and calculate the Schmidt coefficients of
the input states, escaping from the cumbersome task to compute the
evolution equation of every different input state. Another fortunate
thing is that unlike Eq.(8), the RHS of Eq.(14) and Eq.(15) can be
calculated analytically.

Here we would like to point out for a channel $\mathcal {E}$, if
$\tau(({\bf{1}}\otimes\mathcal
{E})|\phi^{+}\rangle\langle\phi^{+}|)=0$, then for arbitrary input
states, we simply find $\tau(({\bf{1}}\otimes\mathcal
{E})|\psi\rangle\langle\psi|)=0$. However, we know
$({\bf{1}}\otimes\mathcal {E})|\psi\rangle\langle\psi|)$ may still
be entangled since $\tau $ is a lower bound of concurrence. In
contrast to concurrence, if $\mathcal {C}[({\bf{1}}\otimes\mathcal
{E})|\phi^{+}\rangle\langle\phi^{+}|]=0$ for a maximally entangled
state, we know for arbitrary input states, $\mathcal
{C}[({\bf{1}}\otimes\mathcal {E})|\psi\rangle\langle\psi|]=0$, the
output states are always separable. We know that $\mathcal {E}$ is
the entanglement breaking channel.

We can generalize Eq.(14) to the case where the input state is a
mixed state $\rho_{0}$. Suppose $\rho_{0}$ has a decomposition that
$\rho_{0}=\sum_{i}p_{i}|\psi_{i}\rangle\langle \psi_{i}|$. By the
convexity of $\tau$, we have $\tau(({\bf{1}}\otimes\mathcal
{E})\rho_{0})=\tau(\sum_{i}p_{i}({\bf{1}}\otimes\mathcal
{E})|\psi_{i}\rangle\langle \psi_{i}|)
\leq\sum_{i}p_{i}\tau(({\bf{1}}\otimes\mathcal
{E})|\psi_{i}\rangle\langle \psi_{i}|)
\leq\frac{d^{2}}{4}\tau(({\bf{1}}\otimes\mathcal
{E})|\phi^{+}\rangle\langle\phi^{+}|)\sum_{i}p_{i}\mathcal
{C}^{2}(|\psi_{i}\rangle)\nonumber$. Considering all decompositions
of $\rho_{0}$, it's easy to see
\begin{eqnarray}
\tau(({\bf{1}}\otimes\mathcal
{E})\rho_{0})\leq\frac{d^{2}}{4}\tau(({\bf{1}}\otimes\mathcal
{E})|\phi^{+}\rangle\langle\phi^{+}|)\tau'(\rho_{0}),
\end{eqnarray}
where $\tau'(\rho_{0})$ is the tangle of $\rho_{0}$ and it has an
easily computable formula for a bipartite mixed state $\rho_{0}$
having no more than two nonzero eigenvalues \footnote{T.J.Osborne,
Phys. Rev. A \textbf{72}, 022309 (2005).} and some states with high
symmetry like isotropic states \footnote{P.Rungta and C.M.Caves,
Phys. Rev. A \textbf{67}, 012307 (2003).}.

In fact $\tau'$ itself has a similar dynamical property to Eq.(14)
and Eq.(16). In the following proof we neglect the normalization
coefficients $p$, $p'$ and $p''$ for simplicity. First we suppose
the input state is pure. If both $\rho'$ and $\rho_{\mathcal {E}}$
are pure states then according to the definition of $\tau'$ and
Eq.(8) we have $\tau'(\rho')\leq\frac{d^{2}}{4}\tau'(\rho_{\mathcal
{E}})\tau'(|\psi\rangle)$. When $\rho_{\mathcal {E}}$ is mixed and
has an optimal decomposition $\rho_{\mathcal
{E}}=\sum_{i}\lambda_{i}|\varphi_{i}\rangle\langle\varphi_{i}|$ such
that $\tau'(\rho_{\mathcal
{E}})=\sum_{i}\lambda_{i}\tau'(|\varphi_{i}\rangle)$, we have
\begin{eqnarray}
\tau'(\rho')
\leq\sum_{i}\lambda_{i}\tau'((\mathcal
{E}_{\psi}\otimes{\bf{1}})|\varphi_{i}\rangle\langle\varphi_{i}|)\nonumber\\
\leq\frac{d^{2}}{4}\sum_{i}\lambda_{i}\tau'(|\varphi_{i}\rangle)\tau'(|\psi\rangle)
=\frac{d^{2}}{4}\tau'(\rho_{\mathcal {E}})\tau'(|\psi\rangle).
\end{eqnarray}
For the case of mixed input state, by the convexity of $\tau'$,
Eq.(17) also holds replacing $|\psi\rangle$ with $\rho_{0}$.

{\label{sec:level1}} \section{Examples and discussion} Suppose
$\mathcal{E}$ is a depolarizing channel, such that $\mathcal
{E}(\rho)=(1-\varepsilon)\rho+\varepsilon\frac{1}{d}{\bf{1}}$ with
$\varepsilon\in[0,1]$. Using the definition in Eq.(4) to decompose
$({\bf{1}}\otimes\mathcal {E})|\psi\rangle\langle\psi|$ into some
$4\times4$ matrices, we find that $\tau(({\bf{1}}\otimes\mathcal
{E})|\psi\rangle\langle\psi|)=\sum_{i<j=0}^{d-1}x_{ij}^{2}$, where
$x_{ij}=\max\{0,\frac{2d-(2d+2)\varepsilon}{d}\sqrt{\omega_{i}\omega_{j}}\}$.
Next we suppose $\mathcal {E}$ is a phase damping channel, namely
$\mathcal
{E}(\rho)=(1-\varepsilon)\rho+\varepsilon\sum_{i=0}^{d-1}\rho_{ii}|i\rangle\langle
i|$ for an input state $\rho$. Through calculation similar to that
of the depolarizing channel, we obtain
$\tau(({\bf{1}}\otimes\mathcal
{E})|\psi\rangle\langle\psi|)=\sum_{i<j=0}^{d-1}y_{ij}^{2}$, where
$y_{ij}=\max\{0,2(1-\varepsilon)\sqrt{\omega_{i}\omega_{j}}\}$.

Now we study the case where the input state is mixed, for example an
isotropic state
$\rho_{F}=\frac{1-F}{d^{2}-1}({\bf{1}}-|\phi^{+}\rangle\langle\phi^{+}|)+F|\phi^{+}\rangle\langle\phi^{+}|$,
where $F=\langle\phi^{+}|\rho_{F}|\phi^{+}\rangle\in[0,1]$. Due to
its invariance under transformation $\mathcal {T}(\rho)=\int
dU(U\otimes U^{\ast})\rho(U\otimes U^{\ast})^{\dagger}$, there exist
elegant formulas for its tangle as well as concurrence [8]. Noting
that if one qudit of $\rho_{F}$ goes through a depolarizing channel,
$\rho_{F}$ is transformed into another isotropic state $\rho_{F'}$
with $F'=F-\frac{Fd^{2}-1}{d^{2}}\varepsilon$ and for isotropic
states $\tau(\rho_{F})$ is exactly equal with $\mathcal
{C}^{2}(\rho_{F})$ [6], we have for $F\leq\frac{1}{d}$,
$\tau(({\bf{1}}\otimes\mathcal {E})\rho_{F})=0$ and for
$F>\frac{1}{d}$, $\tau(({\bf{1}}\otimes\mathcal
{E})\rho_{F})=\frac{2d}{d-1}\Big(\max\Big\{0,F-\frac{1}{d}-\frac{Fd^{2}-1}{d^{2}}\varepsilon\Big\}\Big)^{2}$.

We focus our attention on the dynamics of $\tau$ (see Fig.1 and
Fig.2). For depolarizing channel, we find that when
$\varepsilon\geq\frac{d}{d+1}$, $\tau(({\bf{1}}\otimes\mathcal
{E})|\psi\rangle\langle\psi|)$ vanishes. A similar phenomenon
appears for $\tau(({\bf{1}}\otimes\mathcal {E})\rho_{F})$ when
$\varepsilon\geq\frac{Fd^{2}-d}{Fd^{2}-1}$. This is a sudden death
of $\tau$, similar to the sudden death of entanglement
\footnote{T.Yu and J.H.Eberly, Phys. Rev. Lett. 97, 140403
(2006).}\footnote{M.P.Almeida \emph{et al}, Science \textbf{316},
579 (2007).}\footnote{L.Aolita, R.Chaves, D.Cavalcanti, A.Acin and
L.Davidovich, Phys. Rev. Lett. \textbf{100}, 080501
(2008).}\footnote{C.E.Lopez, G.Romero, F.Lastra, E.Solano and
J.C.Retamal, Phys. Rev. Lett. \textbf{101}, 080503 (2008).}. We hope
$\tau$ will not vanish in a finite time because then it can provide
a non-trivial lower bound to squared concurrence and the state being
evolving is still distillable [6].
\begin{figure}
\includegraphics[height=8cm,width=\linewidth]{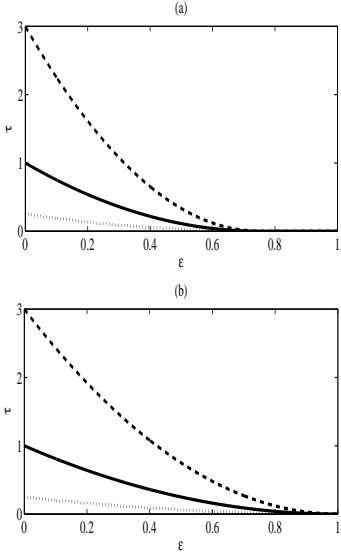}
\caption{\label{fig:epsart} The decay of
$\tau(({\bf{1}}\otimes\mathcal {E})|\psi\rangle\langle\psi|)$(solid
line) and its upper(dashed line) and lower bound(dotted line), where
(a):$\mathcal {E}$ is a depolarizing channel and (b):$\mathcal {E}$
is a phase damping channel. Here we let $d=3$,
$\omega_{0}=\omega_{1}=\frac{1}{6}$ and $\omega_{2}=\frac{2}{3}$.
Note that in (a) a sudden death of $\tau$ appears but in (b) it
doesn't.}
\end{figure}

\begin{figure}
\includegraphics[height=8cm,width=\linewidth]{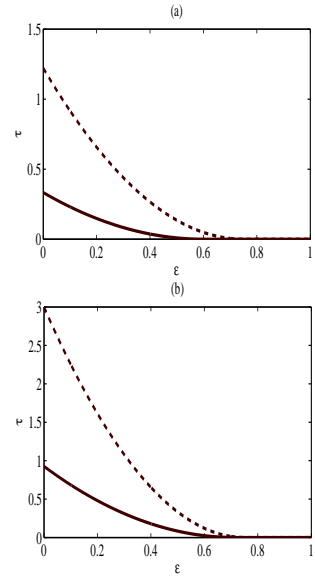}
\caption{\label{fig:epsart} The decay of
$\tau(({\bf{1}}\otimes\mathcal {E})\rho_{F})$(solid line) and its
upper bound(dashed line), where $\mathcal {E}$ is a depolarizing
channel and $d=3$. (a): $F=\frac{2}{3}$; (b): $F=\frac{8}{9}$. Note
that $\tau$ and its upper bound both vanish in finite time, although
 maybe not at the same time.}
\end{figure}
Just like the sudden death of entanglement cannot appear for any
channel [9], the sudden death of $\tau$ doesn't exist for some
channels. For example, for phase damping channel, we can see only
when $\varepsilon=1$, $\tau(({\bf{1}}\otimes\mathcal
{E})|\psi\rangle\langle\psi|)$ vanishes, which means $\tau$ doesn't
die suddenly but asymptotically. But, if the input state is mixed,
for example a $3\otimes3$ Werner state, the sudden death of $\tau$
can also appear even for phase damping channel.

{\label{sec:level1}} {\it Summary.---} The dynamics of a system is a
fundamental feature to describe its time evolution property. In this
paper, we have shown the dynamical properties of the lower and upper
bounds of squared concurrence respectively. Unlike the concurrence
itself, the lower bound of the squared concurrence in this paper is
computable. Thus our results are more reachable in various
situations. We use depolarizing and phase damping channels as
examples and find $\tau$ will vanish in finite time. Whether the
entanglement sudden death appear depends both on the channel and the
input state. Our result provides an easy way to estimate the
dynamics of the entanglement in realistic physical systems.

{\it Acknowledgements}: HF acknowledges the support by "Bairen"
program, NSFC grant (10674162) and "973" program (2006CB921107).
\newpage
\bibliography{apssamp}
\end{document}